\documentclass[entropy,article,accept,pdftex,moreauthors]{Definitions/mdpi}

\firstpage{1} 
\pubvolume{27}
\issuenum{4}
\articlenumber{344}
\pubyear{2025}
\copyrightyear{2025}
\externaleditor{Michael L. Mayo and Kevin R. Pilkiewicz} 
\datereceived{4 February 2025} 
\daterevised{18 March 2025} 
\dateaccepted{19 March 2025 } 
\datepublished{ } 
\hreflink{https://doi.org/} 

\Title{The Geometry of Concepts: Sparse Autoencoder Feature Structure}

\TitleCitation{The Geometry of Concepts: Sparse Autoencoder Feature Structure}

\def\beq#1{\begin{equation}\label{#1}}
\def\eeq{\end{equation}}
\def\beqa#1{\begin{eqnarray}\label{#1}}
\def\eeqa{\end{eqnarray}}

\def\Fig#1{Figure~\ref{#1}}

\usepackage{graphicx}
\usepackage{subcaption}
\usepackage{booktabs}
\usepackage{array}
\usepackage{float}
\usepackage{comment}

\usepackage{bm}
\usepackage[normalem]{ulem}
\usepackage{xcolor}
\usepackage{url}

\newcommand{\change}[1]{\textcolor{black}{#1}}

\Author{Yuxiao 
 Li 
~$^{4,\dagger,\orcidA{}}$, Eric J. Michaud~$^{1,3,\dagger}$, David D. Baek~$^{2,3,\dagger}$, Joshua Engels~$^{2}$, Xiaoqing Sun~$^{1}$ and Max Tegmark
~$^{1,3}$*
}

\AuthorNames{Yuxiao Li, Eric J. Michaud, David D. Baek, Joshua Engels, Xiaoqing Sun and Max Tegmark}

\isAPAStyle{%
       \AuthorCitation{Lastname, F., Lastname, F., \& Lastname, F.}
         }{%
        \isChicagoStyle{%
        \AuthorCitation{Lastname, Firstname, Firstname Lastname, and Firstname Lastname.}
        }{
        \AuthorCitation{Li, Y.; Michaud, E.J.; Baek, D.D.; Engels, J.; Sun, X.; Tegmark, M.}
        }
}

\address{%
$^{1}$ \quad Department of Physics, Massachusetts 
 Institute of Technology, Cambridge, MA, 02139, USA; {ericjm@mit.edu (E.J.M.); xqsun@mit.edu~(X.S.)}
 \\
$^{2}$ \quad Department of Electrical Engineering and Computer Science, Massachusetts Institute of Technology, Cambridge, MA, 02139, USA;
{dbaek@mit.edu (D.D.B.); jengels@mit.edu (J.E.);}
\\
$^{3}$ \quad Institute 
 for Artificial Intelligence and Fundamental Interaction, Cambridge, MA, 02139, USA  
 \\
$^{4}$ \quad Beneficial AI Foundation (BAIF), Cambridge, MA, 02139, USA; {yuxiao2234@gmail.com (Y.L.);}  
}

\corres{Correspondence: tegmark@mit.edu}

\firstnote{These 
 authors 
 contributed equally to this work.}

\abstract{{Sparse autoencoders have recently produced dictionaries of high-dimensional vectors corresponding to the universe of concepts represented by large language models.}
We find that this concept universe has interesting structure at three levels:
(1) The ``atomic'' small-scale structure contains ``crystals'' whose faces are parallelograms or trapezoids, generalizing well-known examples {such as} 
  ({\it man:woman::king:queen}). We find that the quality of such parallelograms and associated function vectors improves greatly when projecting out global distractor directions such as word length, which is efficiently performed with linear discriminant analysis. (2) The ``brain'' intermediate-scale structure has significant spatial modularity; for example, math and code features form a ``lobe'' akin to functional lobes seen in neural fMRI images. We quantify the spatial locality of these lobes with multiple metrics and find that clusters of co-occurring features, at coarse enough scale, also cluster together spatially far more than one would expect if feature geometry were random. (3) The ``galaxy''-scale large-scale structure of the feature point cloud is not isotropic, but instead has a power law of eigenvalues with steepest slope in middle layers. We also quantify how the clustering entropy depends on the layer.}


\begin{document}


\section{Introduction}

While large language models (LLMs) now exhibit a variety of impressive abilities~\cite{hurst2024gpt, TheC3, guo2025deepseek},
we largely do not understand the internal cognition that underlies the behavior of these systems.
{This lack of transparency may pose a challenge for a variety of AI safety~\cite{slattery2024ai} concerns. For~instance, it may be difficult to tell whether seemingly benign model behavior in any particular instance is sycophantic~\cite{sharma2023towards} or deceptive~\cite{park2023ai} without an analysis of the internals of the system. Such ``interpretability'' analysis has already shown promise in auditing AI systems~\cite{marks2024auditing} to identify misaligned goals~\cite{ngo2022alignment}. As~systems become more powerful, there is a need for methods to further our understanding of the internal representations and algorithms learned by these systems~\cite{bereska2024mechanistic,sharkey2025open}.}









The past year has seen a breakthrough in understanding how large language models work:
sparse autoencoders (SAEs) have discovered large numbers of vectors (``features'') in their activation space that can be interpreted as concepts \citep{huben2023sparse,bricken2023towards, templeton2024scaling}. \change{These advances build on earlier studies applying sparse coding to artificial neural network representations~\cite{faruqui2015sparse, zhang2019word,yun2021transformer}, and~to earlier work in neuroscience on biological neural representations~\cite{olshausen1996emergence,olshausen1997sparse}.} Underlying this work is the idea that neural networks use \emph{{sparse}
 coding} to represent concepts in their activation space~\cite{elhage2022superposition}. In~particular, sparse autoencoders are motivated by the assumptions that (1) networks compute a variety of ``features'' from their input, (2) features are represented as one-dimensional directions in activation {space} 
 $\{ \bm{d}_i \}$, (3) features are represented simply by adding them to the network's activations, so activation vectors take the form $\sum_i f_i \bm{d}_i$, and~(4) the coefficients $f_i$ are \emph{sparse}---only a small subset of all possible features ``fire'' at once. The~combination of assumptions (2)--(4) has been called the Linear Representation Hypothesis~\cite{park2023linear, olah2024linear, engels2024not}.

If these assumptions hold, we could automatically discover these features with \emph{{sparse} 
 dictionary learning}. Sparse dictionary learning attempts to learn an overcomplete basis (dictionary) $\{ \bm{d}_i\}$ such that vectors $\bm{x}$ from a given distribution can be represented as sparse linear combinations of dictionary elements. Sparse autoencoders offer a simple approach to sparse dictionary learning. Sparse autoencoders consist of a learnable encoder function $\text{Enc}$, which maps vectors $\bm{x} \in \mathbb{R}^n$ to a hidden latent representation $\bm{f} \in \mathbb{R}^m$, and~a decoder $\text{Dec}$, which maps latent $\bm{f}$ back to $\bm{\hat{x}} \in \mathbb{R}^n$. The~objective of the sparse autoencoder is to accurately reconstruct the {input} 
 $\bm{x}$ from a \textbf{{sparse}} latent representation, and they~are trained with gradient descent with a loss function like
$$ \mathcal{L} = ||\bm{x} - \text{Dec}(\text{Enc}(\bm{x})) ||_2^2 + \lambda ||\bm{f}||_0. $$
{Sparse} 
 autoencoders use a linear decoder $\text{Dec}(\bm{f}) = \bm{W}_d \bm{f} + b_d$, so that the output of the SAE can be interpreted as a linear combination of features: $\bm{\hat{x}} = \sum_i f_i \bm{W}_d^i + \bm{b}_d $. In~practice, hidden latents discovered by sparse autoencoders tend to be more interpretable than neurons, activating in more consistent contexts~\cite{huben2023sparse, bricken2023towards}, suggesting that they may be learning the true latents underlying the network's computation. \change{For AI safety, sparse autoencoders have shown some preliminary success:  Ref.~\cite{marks2024auditing} reports specially training an LLM to have a hidden objective, and~then challenging separate teams of researchers to identify this objective. One team was able to quickly identify this objective by looking at sparse autoencoder features that activated when the LLM was prompted to exhibit ``potentially concerning
behaviors'', and~then looking at examples in the training data where that same feature fired.}

Although some early work motivating sparse autoencoders suggested that networks would arrange features maximally spread apart (approximately orthogonal)~\cite{elhage2022superposition}, recent works have suggested that features may have a more sophisticated geometric structure~\mbox{\cite{engels2024not,templeton2024scaling}}. Recently, a~large collection of SAEs have been made publicly available~\cite{lieberum2024gemma}, so it is timely to study their structure at various scales. Thus, the~present paper examines sparse autoencoder feature structure at three separate spatial scales, \change{which we refer to informally as the ``atom''-scale, ``brain''-scale, and~``galaxy''-scale. These playful analogies are not meant to be precise, but~instead gesture at certain concepts and methods of analysis from other fields which we apply to understanding language model feature structure.} We provide project code at \url{https://github.com/ejmichaud/feature-geometry} (accessed on 24 March 2025).



This paper is organized as follows. In~\Cref{sec:related-work}, we summarize related work. In~\Cref{sec:atoms}, we investigate if the ``atomic'' small-scale structure contains ``crystals'' whose faces are parallelograms or trapezoids, generalizing well-known examples such as { (\it {man:woman::king:queen}}).
In~\Cref{sec:brains}, we test if the ``brain'' intermediate-scale structure has  functional modularity akin to biological brains.
In~\Cref{sec:galaxy}, we study the ``galaxy'' large-scale structure of the feature point cloud, testing whether it is more interestingly shaped and clustered
than an isotropic 
Gaussian distribution, and~conclude in \Cref{sec:conclusion}.

\section{Related~Work}\label{sec:related-work}

\textbf{{Neural} 
 network geometry}: Many past works have studied the geometry of neural network activations. These works find that the intrinsic dimension of neural network hidden states are much lower than the full model dimension~\cite{ansuini2019intrinsic}, that nearby vectors in activation space are semantically similar~\cite{chandrasekaran2021evolution}, and~that at local minima well generalizing neural network loss landscapes have many ``flat'' directions~\cite{watanabe2009algebraic}. Other works study how representations evolve through models; one hypotheses is ``iterative inference'', which claims that neural networks iteratively refine activations layer by layer~\cite{rushing2024explorations, belrose2023eliciting}. A~contrasting hypothesis is a circuits view, which holds that information flows in discrete steps along a directed acyclic graph through the model, and~representations cleanly change between steps~\cite{conmy2023towards}. \change{Another work~\cite{park2024geometry} found that representations of hierarchically related concepts are orthogonal to each other while categorical concepts are represented as polytopes.}  Our work is in the same vein as these earlier analysis, but~differs in an important way because we use the SAE basis, which represents the model's atomic concept space instead of its activation~space.

\textbf{{SAE} 
 feature structure}: Sparse autoencoders (SAEs) are a recent approach for discovering interpretable language model features without supervision, although~relatively few works have examined SAE feature structure. \citet{bricken2023towards} \mbox{and~\citet{templeton2024scaling}} both visualize SAE features with UMAP projections and notice that features tend to group together in ``neighborhoods'' of related features, in~contrast to the approximately orthogonal geometry observed in the toy model of~\citet{elhage2022superposition}. \citet{engels2024not} find examples of SAE structure where multiple SAE features appear to reconstruct a multi-dimensional feature with interesting geometry, and~multiple authors have recently speculated that SAE vectors might contain more important structures \citep{mendel2024geometry, smith2024weak}. \citet{bussmann2024showing} suggest that SAE features are in fact linear combinations of more atomic features, and~discover these more atomic latents with ``meta SAEs''. Our discussion of crystal structure in SAE features is related to this idea that seemingly atomic representations might be composed of more atomic~representations.


\textbf{{Semantically} 
 meaningful linear representations}: Early work found that word embedding methods such as GloVe and Word2vec contained directions encoding semantic concepts, e.g.,~the well-known formula f(king) $-$ f(man) + f(woman) = f(queen)~\citep{drozd2016word, pennington2014glove, ma2015using}. More recent research has found similar evidence of linear representations in sequence models trained only on next token prediction, including Othello board positions~\citep{nanda2023emergent, li2022emergent}, integer lattices \citep{michaud2024opening}, the~truth value of assertions~\citep{marks2023geometry}, and~numeric quantities such as longitude, latitude, birth year, and~death year~\citep{gurnee2023language, heinzerling2024monotonic}, inspiring the Linear Representation Hypothesis (see above). Recent works have also found causal \textit{{function vectors}
} for in-context learning \citep{todd2023function,hendel2023context,kharlapenko2024extracting}. These function vectors induce the model to perform a certain task when added into the model's hidden states. Our discussion of crystal structures builds upon these previous works by finding these task vectors and parallelogram structures in sparse autoencoder representations.


\section{``Atom''-Scale: Crystal~Structure}\label{sec:atoms}

\def\a{{\bf a}}
\def\b{{\bf b}}
\def\c{{\bf c}}
\def\d{{\bf d}}

\begin{figure}[H]
    \includegraphics[width=\linewidth]{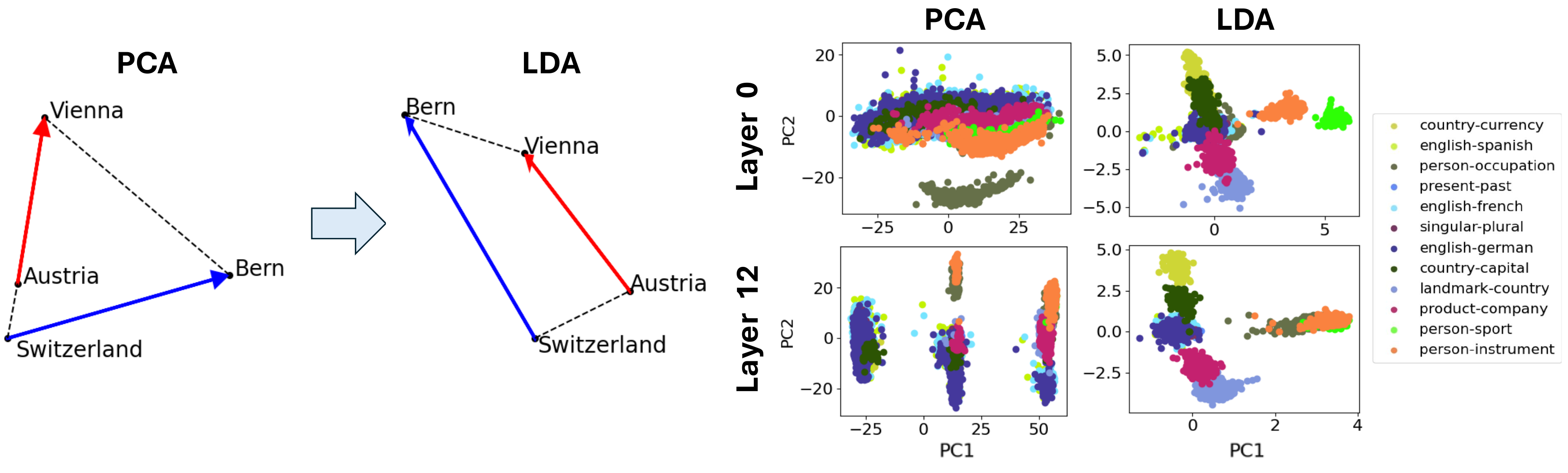}
    \caption{Parallelogram 
 and trapezoid structure is revealed (\textbf{left}) when \change{distractor dimensions were projected out from the activations using LDA. LDA results in
    tighter clusters of 
    pairwise Gemma-2-2b activation differences (\textbf{right}), where each cluster corresponds to a different semantic transformation. Distractor features are defined as those that are not related to semantics of the text; for instance, the~first principal component of Gemma-2-2b's Layer 0 activations (top left figure on the right panel) represents word length. Parallelogram or trapezoid structures suggest that there is a unique direction in the activation space that represents each semantic transformation.}}
    
    \label{fig:lda}
\end{figure}

In this section, we search for what we term  \emph{{crystal}} structure in the point cloud of SAE features. By~this we mean geometric structure
reflecting semantic relations between concepts, 
generalizing the classic example of 
$(\a,\b,\c,\d)$=(man,woman,king,queen) forming an approximate {\it {parallelogram}} where $\b-\a\approx\d-\c$.
This can be interpreted in terms of two {\it {function vectors}}
$\b-\a$ and $\c-\a$ that turn male entities female and 
turn entities royal, respectively.
We also search for {\it {trapezoids}} with only 
one pair of parallel edges $\b-\a\propto\d-\c$ (corresponding to only one function vector); \Fig{fig:lda} (right) shows such an example with $(\a,\b,\c,\d)$=(Austria,Vienna,Switzerland,Bern),
where the function vector can be interpreted as mapping 
countries to their capitals.
\change{Studying these crystal structures is important because they provide insight into how LLMs internally represent semantic operations and relational knowledge. For~instance, function vectors help us assess the extent to which semantic structures within models align with human intuition and language logic, as~explored in the recent literature~\cite{todd2023function, meng2022locating}.}

We search for crystals by computing all pairwise difference
vectors and clustering them \change{ using the K-means algorithm~\cite{ahmed2020k}, where the vectors could be either the original model's hidden state activations (\emph{{model crystal}}) or SAE features' decoder vectors (\emph{{SAE crystal}}). We use Gemma-2-2b for the experiment. If~there is a direction that represents each semantic transformation, we expect each resulting cluster to correspond to each function vector. In~other words,}
any pair of difference vectors in a cluster 
\change{will} form a trapezoid or parallelogram, depending on whether the difference vectors are normalized or not before clustering (or, equivalently, whether we quantify similarity between two difference vectors via
Euclidean distance or cosine similarity).

Our initial search for SAE crystals found mostly noise.
To investigate why, we decided to focus on \change{activations of the model's early layer}, where many SAE features correspond to a single token. Since SAE feature vectors in the early layers are often closely related to the corresponding model activations, we believed that studying the activations of these early layers could help clarify why our initial crystal search primarily found noise. Therefore, we studied
Gemma-2-2b residual stream activations for 
previously reported word $\mapsto$ word
function vectors from the dataset of 
\citep{todd2023function}.
\Fig{fig:lda} illustrates that candidate crystal quadruplets 
are typically far from being parallelograms or trapezoids.
This is consistent with multiple papers pointing out that 
(man,woman,king,queen
) is not an accurate parallelogram~either.

We \change{believe} the reason to be the presence of what we term {\it {distractor features}}. \change{We define distractor features to be the features that are not related to semantics of the text.} For example, we find that the horizontal axis 
in \Fig{fig:lda} (right) corresponds mainly to 
word length (Appendix \ref{app:pcomp-diff}, Figure~\ref{fig:pcomp-diff}), which is semantically irrelevant and
wreaks havoc on the trapezoid (left), since 
``Switzerland'' is much longer than the other words.
\change{However, these distractor features were not always interpretable; in some cases, it was difficult to associate features with any clear linguistic property.}

To eliminate such semantically irrelevant distractor vectors, 
we wish to project the data onto a lower-dimensional
subspace orthogonal to them. 
For the \citep{todd2023function} dataset, we accomplish this with linear discriminant analysis (LDA) \citep{xanthopoulos2013linear},
which projects onto signal-to-noise eigenmodes where 
``signal'' and ``noise'' are defined as the covariance
matrices of inter-cluster variation and intra-cluster
variation, respectively. 
\Fig{fig:lda} illustrates that this dramatically
improves the cluster and trapezoid/parallelogram quality, 
highlighting that distractor features can hide existing~crystals.

\section{``Brain''-Scale: Meso-Scale Modular~Structure}\label{sec:brains}

We now zoom out and look for larger-scale structure. 
In particular, we investigate if \emph{ {functionally}} similar
groups of SAE features (which tend to fire together) are also {\it {geometrically}} similar, forming ``lobes'' in 
the activation space. We refer to this analyis as ``brain''-scale because, in~animal brains, functionally similar groups of neurons also typically cluster together spatially.
For example, Broca's area is involved in speech production,
the auditory cortex processes sound, and~the amygdala is primarily associated with processing emotions.
We are curious whether we can find analogous functional modularity in the SAE feature space. \change{While prior work has qualitatively observed that semantically related features are spatially close via UMAP projections of features~\cite{bricken2023towards,templeton2024scaling}, we aim to more precisely quantify the relationship between functional similarity and spatial similarity.}

We test a variety of methods for automatically discovering such functional ``lobes'' and for quantifying if  they are spatially modular. We define a \textbf{l{obe partition} 
} as a partition of the SAE feature point cloud into $k$ subsets (``lobes'') that are computed \emph{{without positional information}}. Instead, we identify such lobes based on them being \emph{{functionally}} related, specifically, tending to fire together within a~document. 

To automatically identify functional lobes, we first compute a histogram of SAE feature co-occurrences. We take Gemma-2-2b and pass documents from The Pile~\cite{gao2020pile} through it. In~this section, we report results with a Layer 12 residual stream SAE with 16k features and 
an average L0 of 41.
For this SAE, we record the features that fire (we count a feature $i$ as firing if its encoder assigns it a coefficient $ f_i> 1$). Features are counted as co-occurring if they both fire within the same block of 256 tokens---this length provides a coarse ``time resolution'' allowing us to find tokens that tend to fire together within the same document rather than just at the same token. We use a max context length of 1024, and~only use one such context per document, giving us at most four blocks (and histogram updates) per document of The Pile. We compute histograms across 50k documents. Given this histogram, we compute an affinity score between each pair of SAE features based on their co-occurrence statistics and perform spectral clustering on the resulting affinity matrix. \change{We use the spectral clustering implementation of scikit-learn~\cite{scikit-learn} with default settings with varying choice of \texttt{n\_clusters}.}

\begin{figure}[H]
    \includegraphics{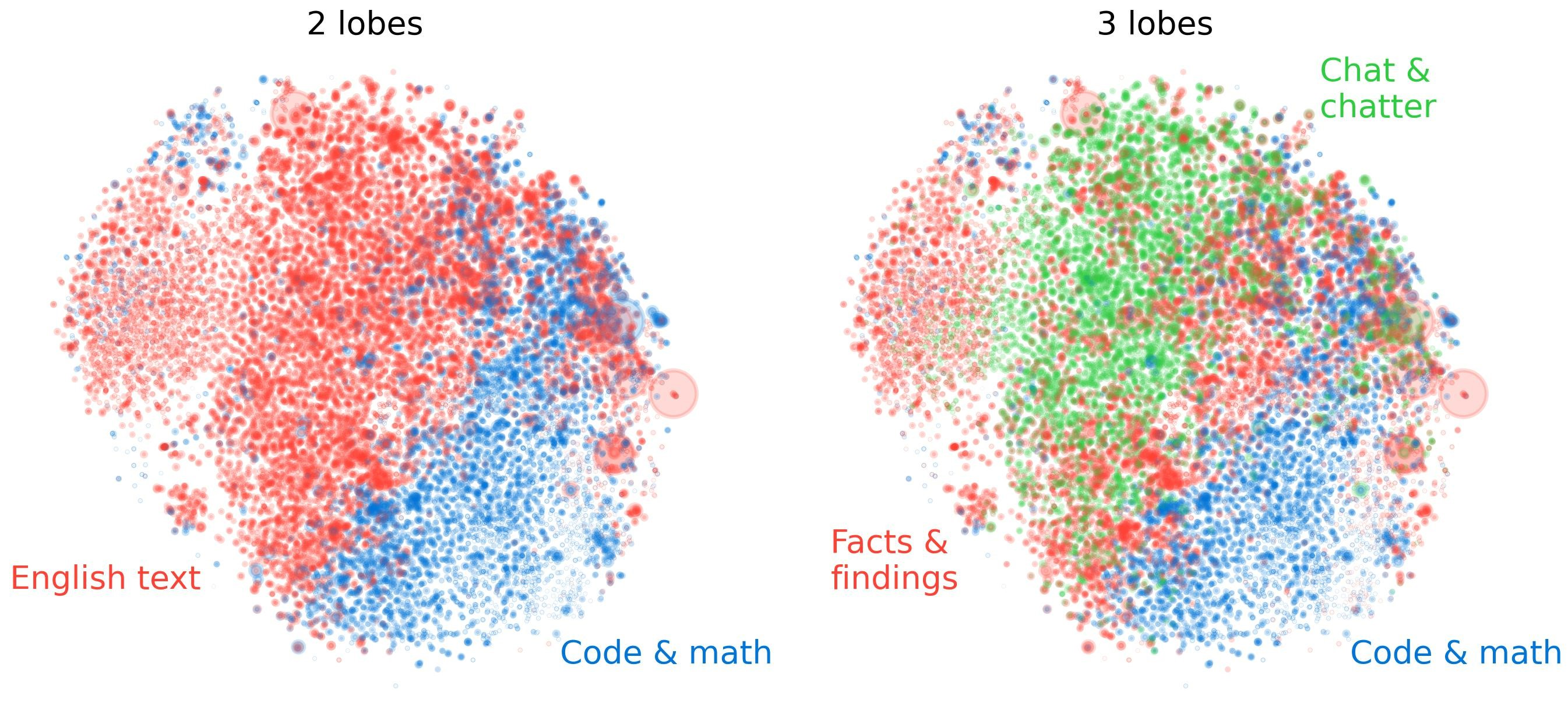}
    \caption{Features 
 in the SAE feature point cloud identified that tend to fire together within documents are seen to also be geometrically co-located in functional ``lobes'', here down-projected to 2D with t-SNE with point size proportional to feature frequency.
    A 2-lobe partition (\textbf{left}) is seen to break the point cloud into roughly 
    equal parts, active on code/math documents
    and English language documents, respectively.
    A 3-lobe partition (\textbf{right}) is seen to mainly subdivide the
    English lobe into a part for short messages and dialogue (e.g., chat rooms and parliament proceedings) and one 
    primarily containing long-form scientific papers.}
    \label{fig:annotated-lobes}
\end{figure}

\change{In \Fig{fig:annotated-lobes}, we visualize lobes discovered with this method with \texttt{n\_clusters}=2,~3 via a t-SNE projection~\cite{van2008visualizing}. For~this figure, we used the ``phi coefficent'' as the measure of co-occurrence similarity between features. We find that lobes visually appear to be spatially localized. For~instance, features which fire primarily on math and code documents tend to cluster together spatially.}

\begin{figure}[H]
    \includegraphics[width=0.8\textwidth]{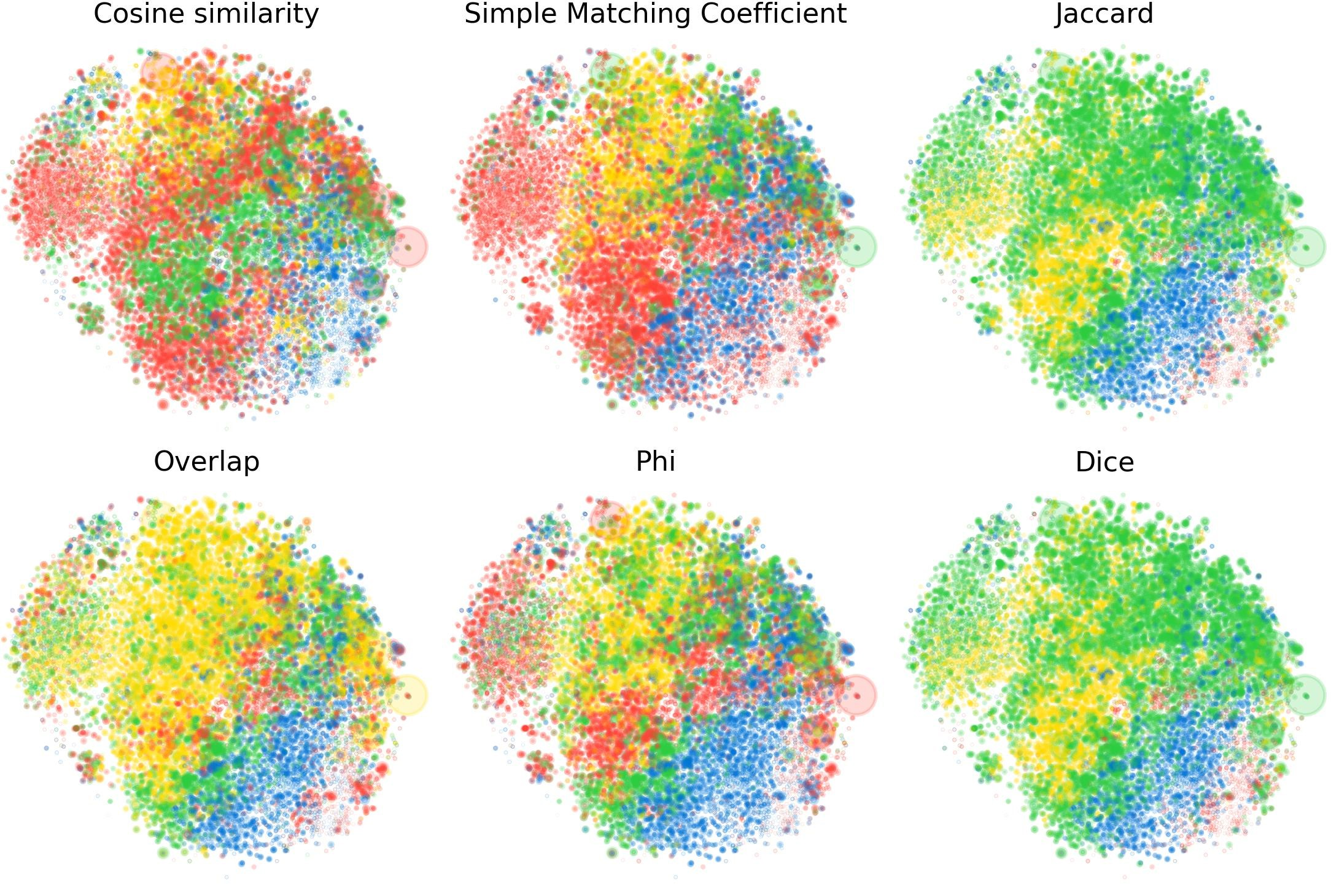}
    \caption{Comparison 
 of the lobe partitions of the SAE point cloud discovered with different affinity measures, with~the same t-SNE projection as Figure~\ref{fig:annotated-lobes}. In~the top left, we show clusters computed from \textbf{{geometry}}, the~cosine similarity between features as the affinity score for spectral clustering. All other measures are based on whether SAE features co-occur (fire together) within 256-token blocks, using different measures of affinity. Although~the phi coefficient predicts spatial structure best, all co-occurrence measures are seen to discover the code/math~lobe.}
    \label{fig:lobe-comparisons}
\end{figure}

We experiment with the following notions of co-occurrence-based affinity: simple matching coefficient, Jaccard similarity, Dice coefficient, overlap coefficient, and~phi coefficient, which can all be computed just from a co-occurrence histogram. In~the Appendix~\ref{app:co-occurrence-measures}, we review definitions for each of these and in Figure~\ref{fig:lobe-comparisons} illustrate how the choice between them
affects the resulting lobe t-SNE plots. \change{We also show how lobes appear when we cluster based on geometry directly using cosine similarities, as~described below.}

\vspace{-6pt}

\change{While these plots show a qualitative relationship between co-occurrence and feature geometry, we aim to quantify this relationship.} Our null hypothesis is that functionally similar points (of commonly co-occurring SAE features) are uniformly distributed throughout the activation space, showing no spatial modularity.
To quantify how statistically significant this is, 
we use two approaches to rule out the null hypothesis:

\begin{enumerate}
    \item While we can cluster features based on whether they co-occur, we can also perform spectral clustering based on the cosine similarity between SAE feature decoder vectors. \change{So instead of feature affinity values being, e.g.,~their co-occurrence phi coefficient, affinity matrix values are instead computed simply from feature geometry as $A_{ij} = \bm{d}_i \cdot \bm{d}_j$.} Given a clustering of SAE features using cosine similarity and a clustering using co-occurrence, we compute the mutual information between these two sets of labels. In~some sense, this measures the amount of information about geometric structure that one obtains from knowing functional structure. We report the adjusted mutual information~\cite{vinh2009information} as implemented by scikit-learn~\cite{scikit-learn}, which corrects for chance agreements between the clusters.
    \item Another conceptually simple approach is to train models to predict which functional lobe a feature is in from its geometry. To~accomplish this, we take a given set of lobe labels from our co-occurrence-based clustering, and~train a logistic regression model to predict these labels directly from the point positions, using 
    an 80-20 train--test split and reporting the balanced test accuracy of this classifier.
\end{enumerate}

\begin{figure}[H]
 \hspace{-5pt}
   \includegraphics{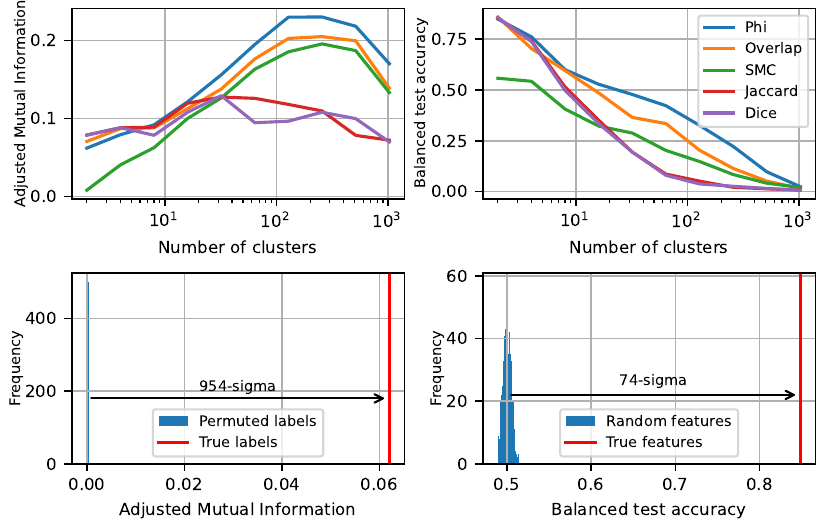}
    \caption{
    (\textbf{top left}): Adjusted 
 mutual information between spatial clusters and functional (co-occurrence-based) clusters. (\textbf{top right}): logistic regression  balanced test accuracy, predicting co-occurrence-based cluster label from position. (\textbf{bottom left}): Adjusted mutual information with randomly permuted cosine similarity-based clustering labels. (\textbf{bottom right}): balanced test accuracy with random unit-norm feature vectors. 
    The statistical significance reported is for phi-based clustering into lobes.}
    \label{fig:affinity-comparisons-four}
\end{figure}

Figure~\ref{fig:affinity-comparisons-four} shows that for both measures, the~phi coefficient gives the best correspondence between functional lobes and feature geometry. To~show that this is statistically significant, we randomly permute the cluster labels from the cosine similarity-based clustering and measure the adjusted mutual information. We also re-initialize the SAE feature decoder directions from a random Gaussian and normalize, and~then train logistic regression models to predict functional lobe from these random feature directions. 
Figure~\ref{fig:affinity-comparisons-four} (bottom
) shows that both tests rule out the null hypothesis with high significance, at~954 and 74 standard deviations, 
respectively, clearly demonstrating that the lobes we see are real and not a statistical~fluke.

\begin{figure}[H]
    \includegraphics[width=0.9\textwidth]{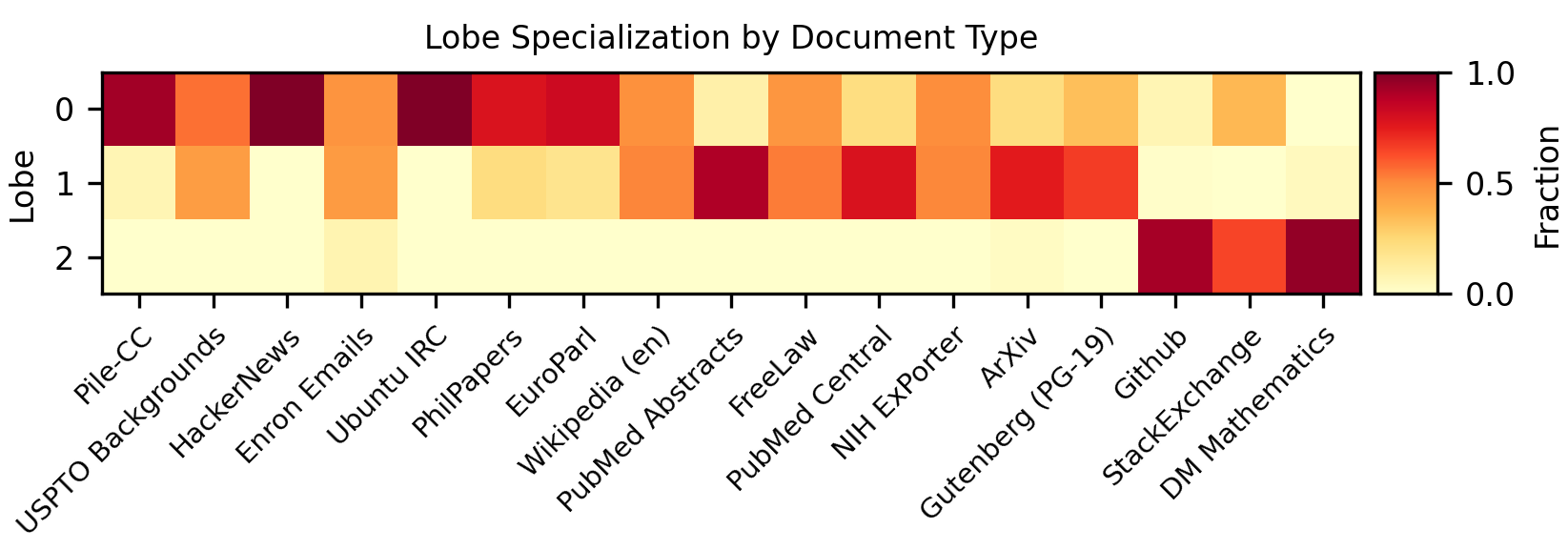}
    \caption{Fraction of contexts in which each lobe had the highest proportion of activating features. For~each document type, these fractions sum to 1 across the lobes. We see that lobe 2 typically disproportionately activates on code and math documents. Lobe 0 and 1 activate on other documents, with~lobe 0 activating more on documents containing short text and dialogue (chat comments, parliamentary proceedings) and lobe 1 activating more on scientific~papers.}
    \label{fig:heatmap3lobes}
\end{figure}

To assess what each lobe specializes in, we run 10k documents from The Pile through Gemma-2-2b, and~again record which SAE features at Layer 12 fire within blocks of 256~tokens. For~each block of tokens, we record which lobe has the highest proportion of its features firing. Each document in The Pile is attached with a name specifying the subset of the corpus that document is from. For~each document type, for~each 256-token block within a document of that type, we record which lobe had the highest proportion of its SAE features firing. Across thousands of documents, we can then look at a histogram of which lobes were maximally activating across each document type. We show these results for three lobes, computed with the phi coefficient as the co-occurrence measure, in~Figure~\ref{fig:heatmap3lobes}. 
This forms the basis for our lobe labeling in Figure~\ref{fig:annotated-lobes}.

These findings raise interesting questions about whether individual sparse autoencoder features are the most natural units for understanding neural networks~\cite{mueller2024quest, olah2023interpretability}. In~biological brains, one can study individual neurons, groups of neurons, groups of groups of neurons, and~so on up to very large-scale structures, and~it is not clear a priori what ``scale'' of analysis will be most fruitful~\cite{hoel2013quantifying}. We may face a similar ambiguity with sparse autoencoder features, since, as~we have seen, groups of co-occurring, geometrically related features can be interpretable and studied in their own right. This question, of~whether there is a right ``scale'' of analysis for SAE features, is made even more salient by the observation in prior work of ``feature splitting''~\cite{bricken2023towards}.

\vspace{-5pt}

\section{``Galaxy''-Scale: Large-Scale Point Cloud~Structure}\label{sec:galaxy}

In this section, we further broaden our perspective to analyze the ``galaxy''-scale structure of the point cloud, focusing on its overall shape and clustering properties. \change{This analysis is loosely inspired by work in astronomy~\cite{kennicutt1998star} characterizing the shape~\cite{hubblenebulae} and substructure~\cite{kravtsov2010dark}
of galaxies.} \change{We start by formulating a simple null hypothesis:}
\change{\textit{{The point cloud is drawn from an isotropic multivariate Gaussian distribution.} 
}}

\begin{figure}[H]
    \includegraphics[width=0.65\linewidth]{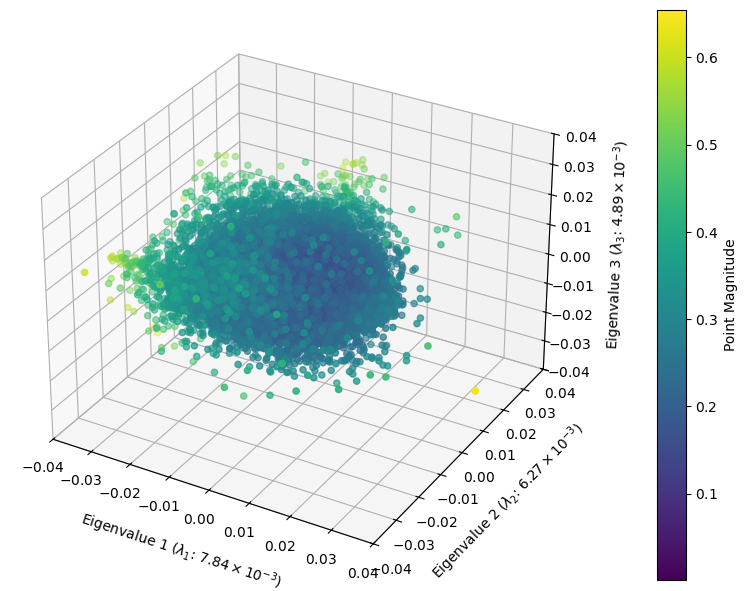}
    \caption{Three-dimensional 
 \textls[-25]{point cloud visualizations of top PCA components for the Gemma-
2-2b Layer 12~SAE~features}.}
    \label{fig:pointcloud_eigenvalues}
\end{figure}

\change{To test this, we analyze the covariance of the data.}
As illustrated in \Fig{fig:pointcloud_eigenvalues}, the~eigenvalue spectrum deviates from isotropy, meaning the cloud exhibits directional structure rather than being purely spherical. \change{Even within the first three principlal components, the~point cloud is anisotrophic,}
with some principal axes slightly wider than~others.

\change{To quantify these deviations, we analyze the eigenvalue spectrum of the covariance matrix, comparing it  to theoretical expectations from random matrix theory (RMT).}

\subsection{Shape~Analysis}

\change{In RMT, the~covariance matrix of $N$ random vectors from a multivariate Gaussian distribution follow a Wishart distribution~\cite{wishart1928generalised}. Under~this assumption, we would expect the eigenvalues to be relatively uniform or to follow the Marcenko--Pastur law~\cite{marchenko1967distribution}. In~contrast, we observe a surprising derivation:}
\begin{itemize}
    \item \change{The eigenvalue spectrum of the point cloud decays as a power law rather than following the expected Wishart behavior.}
    \item \change{As shown in \Fig{fig:pointcloud_eigenvalues}, this power law decay is more pronounced in SAE features compared to raw activations.}
\end{itemize}

Since the abrupt drop off seen for the smallest eigenvalues is caused by limited data and vanishes in the limit $N\to\infty$, we dimensionally reduce the point cloud to its 100 largest principal components for all subsequent analysis in this section.
\change{We describe the shape of this high-dimensional point cloud as resembling a ''fractal cucumber'',}
whose width in successive dimensions falls off like a power law. We find such power law scaling is significantly less prominent for activations than for SAE features; it will be interesting for further work to investigate its~origins.





\begin{figure}[H]
    \includegraphics[width=0.45\linewidth]{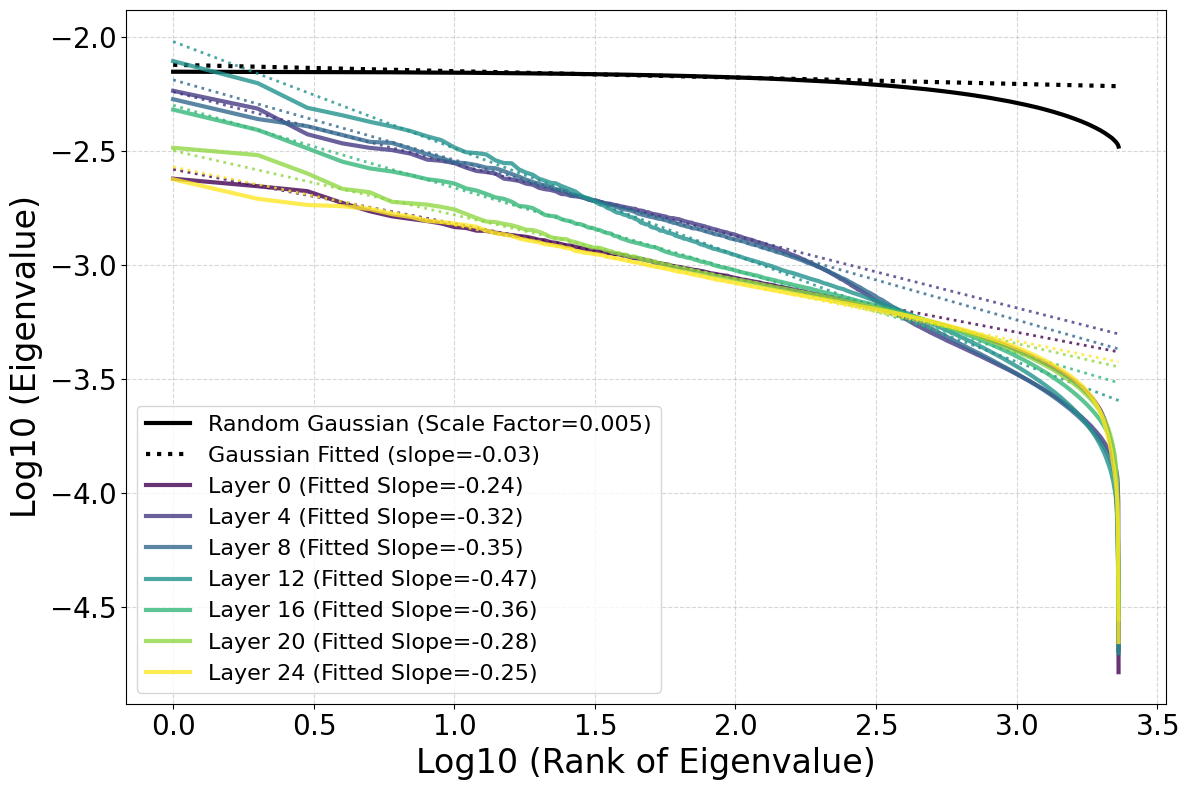}
    \includegraphics[width=0.45\linewidth]{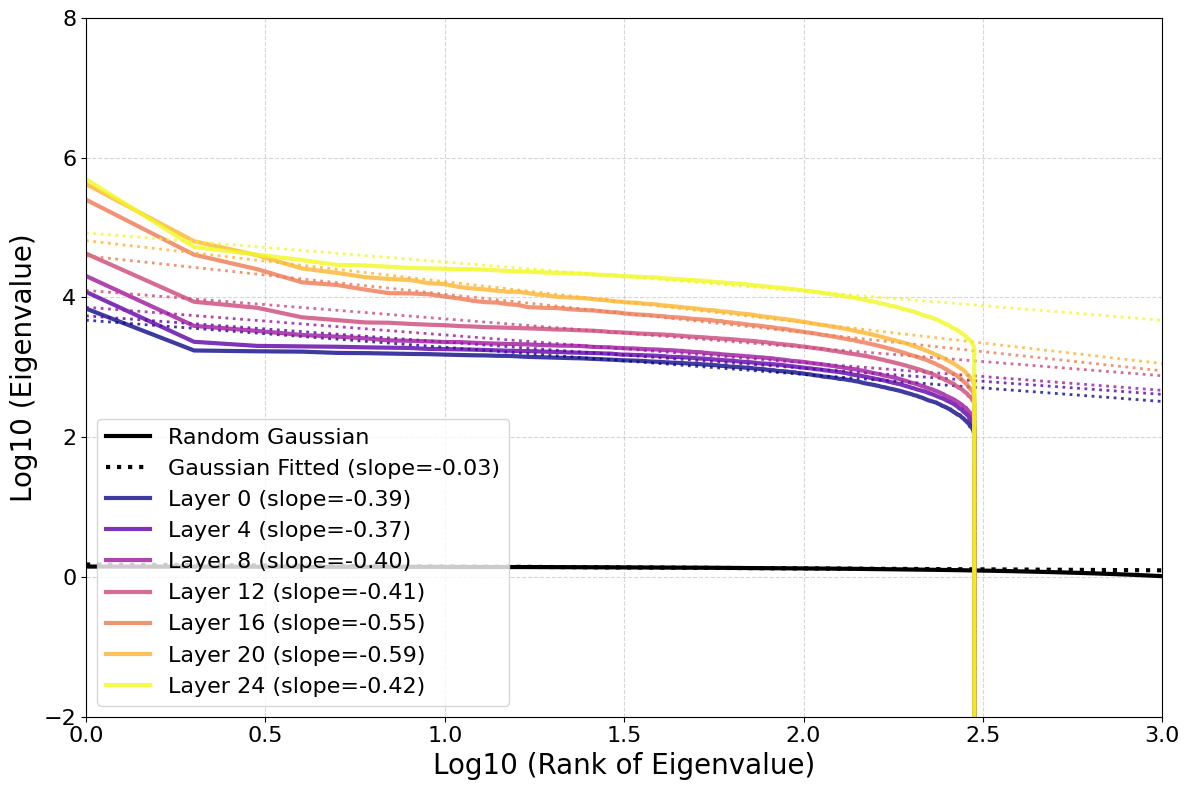}
    \caption{Eigenvalue 
 distributions 
 for SAE features and activations. Eigenvalues of the covariance matrix for SAE features (\textbf{left}) and neural activations (\textbf{right}) decay approximately as a power law, with~slopes varying across layers. A~scaled isotropic Gaussian spectrum is shown for comparison, highlighting the significantly steeper decay for SAE features. Eigenvalue spectra for activations show a much slower decay compared to SAE features, indicating weaker power law   behavior and distinct geometric structures.
    }
    \label{fig:eigenvalues}
\end{figure}
\begin{figure}[H]
    \centering
    \includegraphics[width=0.5\linewidth]{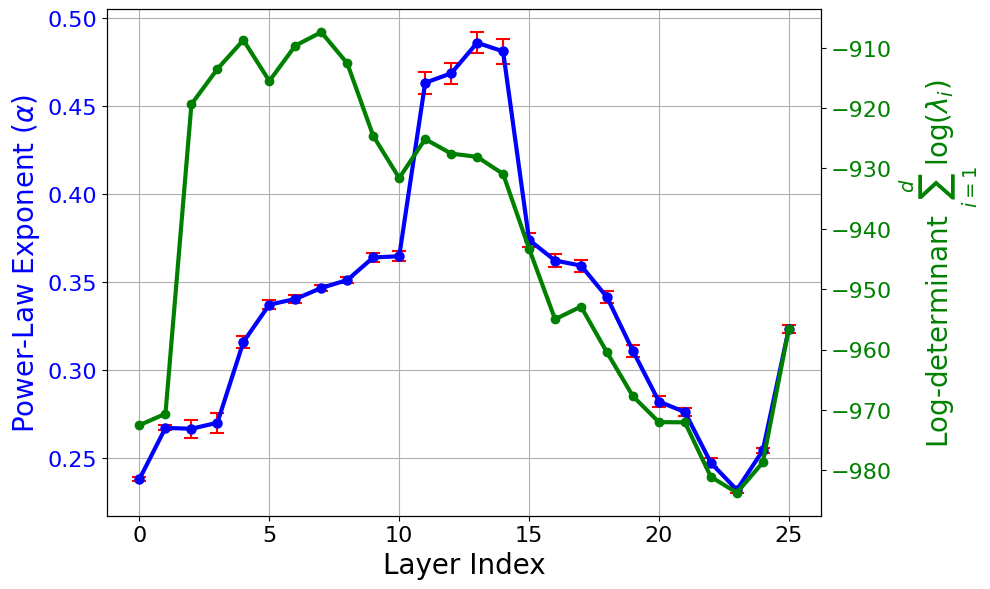}
    \includegraphics[width=0.455\linewidth]{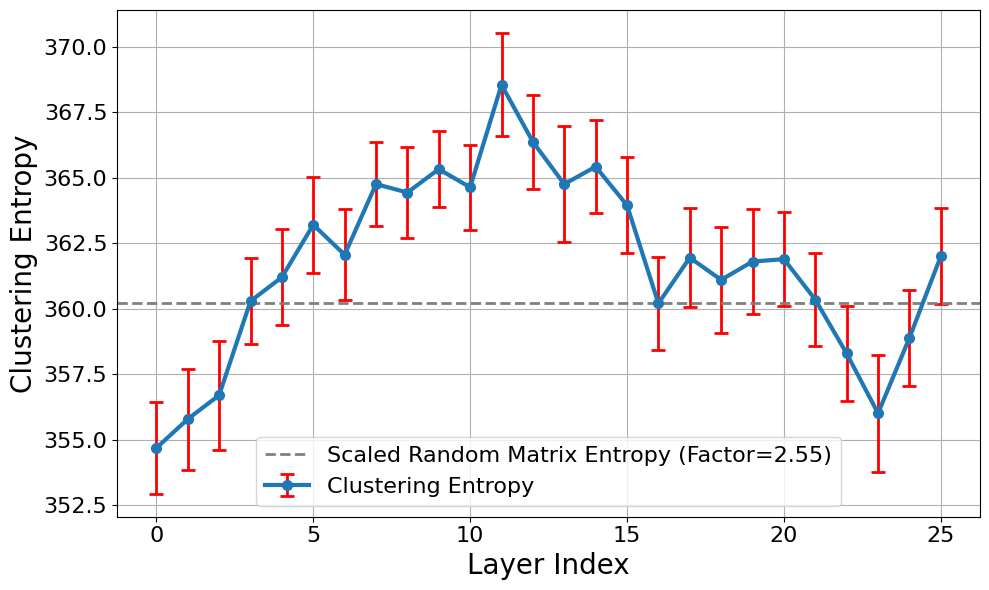}
    \caption{Layer
-wise analysis of latent representations. (\textbf{left}): The power law slope ($\alpha$) of the eigenvalue spectrum (blue) and the log-determinant of the covariance matrix (green) vary across layers. Both metrics peak in intermediate layers, indicating significant structural changes in the latent space.
    (\textbf{right}): Estimated clustering entropy across layers with 95\% confidence intervals. Middle layers exhibit reduced clustering entropy, while earlier and later layers show higher entropy, reflecting distributed and concentrated feature representations, respectively.
    }
    \label{fig:entropy}
\end{figure}

Figure~\ref{fig:eigenvalues} (left) shows how the slope of the 
aforementioned power law depends on LLM layer, computed via linear regression
against the 100 largest eigenvalues.
We see a clear pattern where middle layers have the steepest power law slopes:
(Layer 12 has slope $-$0.47, while early and late layers (e.g., Layers 0 and 24)
have shallower slopes ($-$0.24 and $-$0.25), respectively.
This may hint that middle layers act as a bottleneck, compressing information into fewer principal components, perhaps optimizing for more efficient representation of high-level abstractions.
%
Figure~\ref{fig:eigenvalues} (right) compares the eigenvalue spectra of SAE features and neural activations, indicating a significantly steeper power law decay for SAE features. Activations, in~contrast, exhibit a much slower decay, indicating weaker power law behavior and distinct geometric structures in the latent space.
Figure~\ref{fig:entropy} (left
) explores the effective cloud volume (the determinant of the covariance matrix) of the point cloud, quantified by the log-determinant of the covariance matrix across layer. This volume variation further reflects the layer-specific changes in the structure and complexity of the latent~space.

\vspace{-6pt}

\subsection{Clustering~Analysis}

Clustering of galaxies or microscopic particles is often quantified in terms of a power spectrum or correlation function. This is complicated for our very
high-dimensional data, since the underlying density varies with radius
and, for~a high-dimensional Gaussian distribution, is strongly concentrated around a relatively thin spherical shell.
For this reason, we \change{instead} quantify clustering by estimating the 
{\it {entropy}} of the distribution that the point cloud is assumed to be sampled from.
We estimate the entropy $H$ from our SAE feature point cloud using the $k$-th nearest neighbor (k-NN) method~\cite{dasarathy1991nearest,kozachenko1987sample}, computed as follows,
\begin{equation}
    H_{features} =\frac{d}{n}\sum_{i=1}^n\log(r_i + \theta) 
    + \log(n-1) - \Psi
\end{equation}
where $r_i$ is the distance to the $k$-th nearest neighbor for point $i$, and~$d$ is the dimensionality of the point cloud;
$n$ is the number of points; the constant $\Psi$ is the digamma term from the k-NN estimation.
As a baseline, the~Gaussian entropy represents the maximum possible entropy for a given covariance matrix. For~a Gaussian distribution with the same covariance matrix, the~entropy is computed as follows:
\begin{equation}
    H_{gauss} = \frac{d}{2}\big(1 + \log(2\pi)\big) + \sum_{i=1}^d\log(\lambda_i)
\end{equation}
where $\lambda_i$ are the eigenvalues of the covariance matrix.
We define the \textbf{{clustering entropy}} (often referred to as ``negentropy'' in physics as $H_{gauss}-H$, i.e.,~how much lower the entropy is than its maximum allowed value). 

The estimated clustering entropy is shown in Figure~\ref{fig:entropy} (right
), plotted across different layers. The~results indicate that the SAE point cloud is strongly clustered, particulary in 
the middle layers.
This observation aligns with the reduced clustering entropy seen at intermediate layers, suggesting significant structural differences in the latent~representations.

In future work, it will be interesting to investigate whether these variations depend mainly on the prominence of crystals or lobes in different layers, or~have an altogether  different origin (entirely different underlying mechanisms).

\section{Conclusions}\label{sec:conclusion}
We have \change{searched for structure in the SAE concept universe at three levels:}
(1) The ``atomic'' small-scale structure contains ``crystals'' whose faces are parallelograms or trapezoids, generalizing well-known examples such as ({\it {man:woman::king:queen}}), may be revealed when projecting out semantically irrelevant distractor features.
(2) The ``brain'' intermediate-scale structure has significant spatial modularity; for example, math and code features form a ``lobe'' akin to functional lobes seen in neural fMRI images.
(3) The ``galaxy'' large-scale structure of the feature point cloud is not isotropic, but~instead has a power law of eigenvalues with steepest slope in middle~layers.

\change{While we have observed that SAE features exhibit geometric structure at multiple scales, we have not explained \emph{{why}} this structure forms. We think that further work that not only studies the structure of SAE features, but~also seeks to explain the origin of this structure, could be highly valuable. Such work may lead to refinements to our theory of how networks represent features in superposition or to insights that improve sparse autoencoder performance.}

We hope that our findings serve as a stepping stone toward deeper understanding of SAE features and the workings of large language models, \change{and that this deeper understanding will eventually help to improve the safety of AI systems as they continue to grow in power.}

\vspace{6pt}

\authorcontributions{Conceptualization, J.E. and M.T.; software, E.J.M., D.D.B. and X.S.; formal analysis, Y.L., E.J.M. and D.D.B.; investigation, M.T.; writing---original draft preparation, Y.L., E.J.M., D.D.B., J.E., X.S. and M.T.; writing---review and editing, Y.L., E.J.M., D.D.B., J.E. and M.T.; visualization, Y.L., E.J.M. and D.D.B. All authors have read and agreed to the published version of the manuscript. 
}

\funding{E.J.M., D.D.B, and M.T. are supported by IAIFI through NSF grant PHY-2019786. E.J.M. and J.E. are supported through the NSF GRFP (Grant No. 2141064). This work is supported by
the Rothberg Family Fund for Cognitive Science. 
}

\institutionalreview{Not applicable. 
}

\dataavailability{We provide code to replicate our results at this repository: \url{https://github.com/ejmichaud/feature-geometry} (accessed on 24 March 2025). 
} 

\conflictsofinterest{The authors declare no conflicts of interest. 
} 


\appendixtitles{yes} 
\appendixstart
\appendix

\section{Additional Information on Brain~Lobes}
\unskip

\subsection{Co-Occurrence~Measures}
\label{app:co-occurrence-measures}

Definitions of co-occurrence-based affinity measures: Let $n_{ij}$ be the number of times features $i$ and $j$ co-occur. Let $m_{11}$ be number of times $i$ and $j$ co-occur, $m_{00}$ be number of times $i$ and $j$ both do not occur, $m_{10}$ be number of times $i$ occurs but $j$ does not, $m_{1\bullet}$ be number of times $i$ occurs and $j$ either occurs or not, and~so on. Then, the following can be determined.

Jaccard similarity,~Ref. \cite{jaccard}, is as follows:
$$
J_{ij} = \frac{|i \cap j|}{|i \cup j|} = \frac{n_{ij}}{n_{ii} + n_{jj} - n_{ij}}
$$

Dice score,~Ref. \cite{dicecoeff},is as follows: 
$$
DSC_{ij} = \frac{2|i \cap j|}{|i|+|j|} = \frac{2n_{ij}}{n_{ii}+n_{jj}}
$$

The overlap coefficient is as follows:
$$
overlap_{ij} = \frac{|i \cap j|}{\min{(|i|,|j|)}} = \frac{n_{ij}}{\min{(n_{ii},n_{jj})}}
$$

The simple matching coefficient is 
$$
SMC_{ij} = \frac{m_{00} + m_{11}}{m_{00} + m_{11} + m_{01} + m_{10}}
$$

The phi coefficient,~Ref. \cite{phi_coeff}, is
$$
\phi_{ij} = \frac{m_{11}m_{00} - m_{10}m_{01}}{\sqrt{m_{1\bullet}m_{0\bullet}m_{\bullet1}m_{\bullet0}}}
$$
\vspace{-6pt}


\vspace{-11pt}

\begin{figure}[H]
\hspace{-5pt}        \includegraphics[width=.4\textwidth]{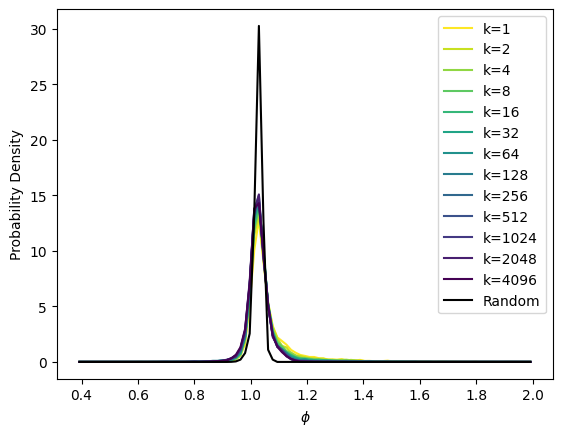}
 ~~~~~~~~~~~~~~~
        \includegraphics[width=.4\textwidth]{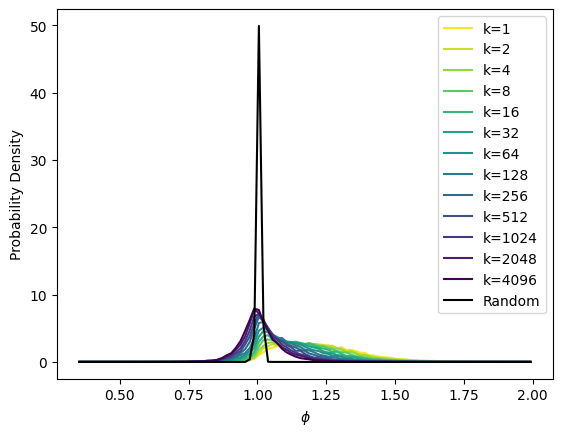}
    \caption{Histogram, over~all features, of~phi coefficient with $k$-th nearest cosine similarity neighbor for (\textbf{left}) Layer 0 and (\textbf{right}) Layer 12. Each line represents a different $k$. The~``random'' line is plotted by drawing a random feature for each feature, then computing the phi coefficient. Features with higher cosine similarity have higher phi coefficients, but~this is less pronounced in Layer 0 compared to Layer~12.}
\end{figure}
\unskip


\section{Understanding Principal Components in Difference~Space}
\label{app:pcomp-diff}

\begin{figure}[H]
    \includegraphics[width=0.55\linewidth]{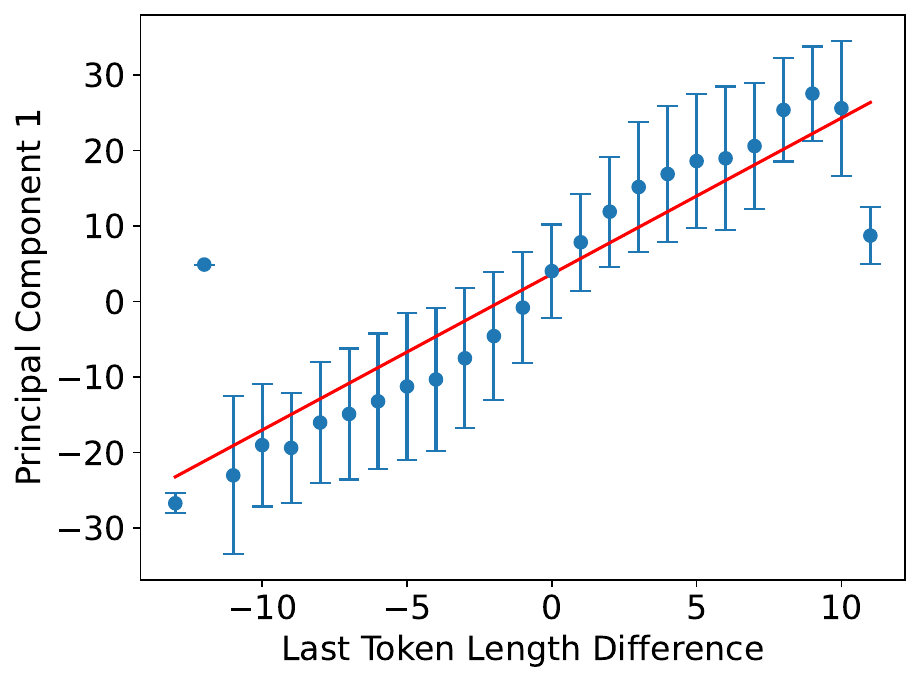}
    \caption{Plot of the first principal component in the difference space as a function of last token length difference in Gemma-2-2b Layer 0. The~linear relationship indicates that the first principal component encodes the length difference between two words' last~tokens.}
    \label{fig:pcomp-diff}
\end{figure}

Figure~\ref{fig:pcomp-diff} shows that the first principal component encodes mainly the length difference between two words' last tokens in Gemma-2-2b Layer~0.

\unskip

\begin{figure}[H]
    \includegraphics[width=0.55\linewidth]{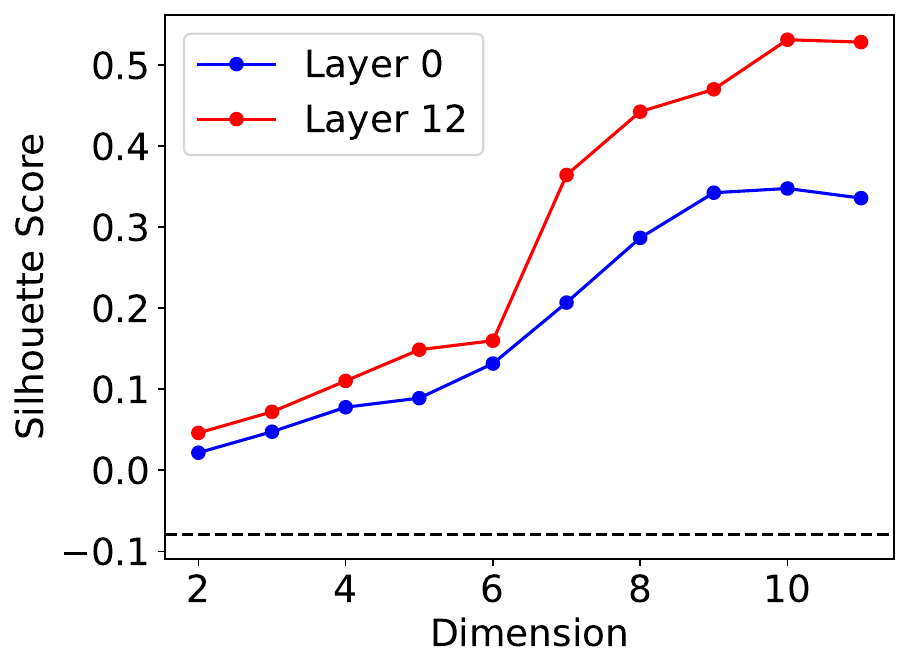}
    \caption{Silhouette score, a~measure of clustering quality, as~a function of reduced dimension in LDA. The~plot indicates that training an affine transformation for semantic cluster separation is easier in middle layers (Layer 12), where the model starts to develop concept-level understanding of the~input.}
    \label{fig:lda-sil}
\end{figure}
\unskip

\section{Breaking Down SAE Vectors by PCA~Component}
\begin{figure}[H]
    \includegraphics[width=0.8\linewidth]{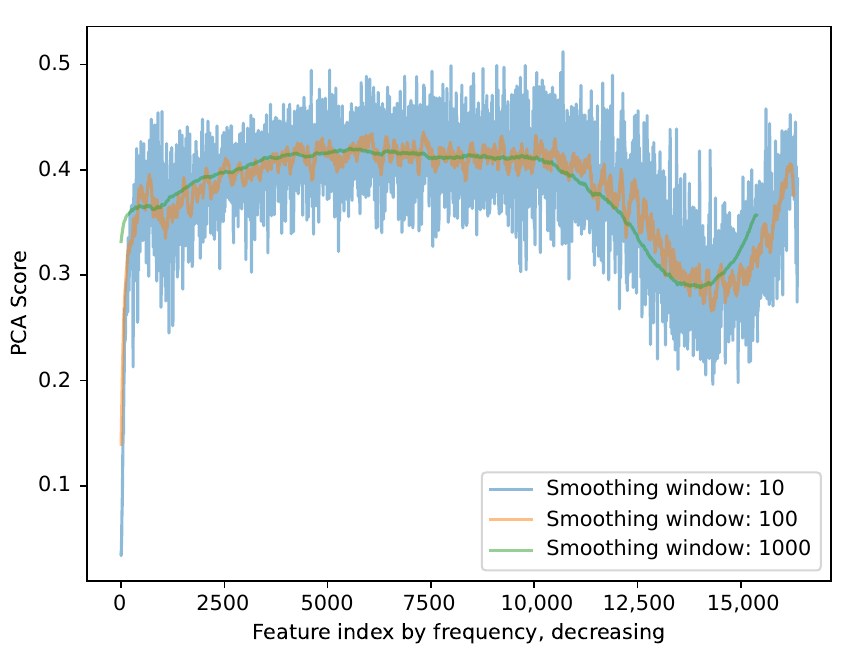}
    \caption{Smoothed 
 PCA scores for each SAE feature of Layer $12$, width 16k 
, $L_0 = 176$ Gemma Scope 2b SAE, sorted by frequency. PCA score = $\frac{1}{n}\sum_{i} i * (pca_i @ feature_j)^2$, where $n$ is the number of PCA features. The~smoothed curves just average this somewhat noisy metric over adjacent sorted features. This measures approximately where in the PCA each SAE feature lies, and~shows that there is a dip into earlier PCA features in the last third of SAE~features.}
    \label{fig:sae-per-pca}
\end{figure}

An additional investigation of structure we undertake is quantifying how SAE vectors are distributed throughout the PCA components of the activations vectors. To~accomplish this, we define a PCA score:
$$
\texttt{PCA score}(feature_j) = \frac{1}{n}\sum_{i} i * (pca_i @ feature_j)^2
$$
{This} 
 metric is a weighted sum between $0$ and $1$ measuring approximately where in the PCA each SAE feature lies. In~Figure~\ref{fig:sae-per-pca}, we plot this metric on a single Gemma Scope SAE (the results look similar on all Gemma Scope SAEs), and~we see that there is an intriguing dip into earlier PCA features in the last third of SAE~features.

\vspace{-4pt}

\begin{adjustwidth}{-\extralength}{0cm}

\reftitle{References}

%


\PublishersNote{}
\end{adjustwidth}
\end{document}